\documentclass[sigconf]{acmart}

\usepackage[utf8]{inputenc}
\usepackage[T1]{fontenc}
\usepackage{color}
\usepackage[caption=false]{subfig}
\usepackage{multirow}
\usepackage{balance}
\usepackage{xspace}
\usepackage{float}
\usepackage{graphicx}
\usepackage{url}
\usepackage[inline]{enumitem} 
\usepackage{enumerate}
\usepackage{booktabs}

\usepackage{mathtools}

\usepackage[printonlyused]{acronym}
\acrodef{SGX}[SGX]{Intel's Software Guard Extensions}
\acrodef{EPC}{Enclave Page Cache}
\acrodef{EPCM}{Enclave Page Cache Map}
\acrodef{MMU}{Memory Management Unit}
\acrodef{AEX}{Asynchronous Enclave Exit}
\acrodef{AEP}{Asynchronous Exit Pointer}
\acrodef{ISR}{Interrupt Service Routine}
\acrodef{OS}{operating system}
\acrodef{DoS}{denial-of-service}
\acrodef{SDK}{Software Development Kit}
\acrodef{TCB}{trusted computing base}
\acrodef{EDL}{Enclave Description Language}
\acrodef{API}{application programming interface}
\acrodef{TEE}{Trusted Execution Environment}
\acrodefplural{TEE}{trusted execution environments}
\acrodef{IAS}{Intel Attestation Service}

\usepackage{todonotes}
\presetkeys%
    {todonotes}%
    {inline}{}

\usepackage{amsmath}
\allowdisplaybreaks[2]

\usepackage{amssymb}

\newcommand{\str}[1]{\textsc{#1}}
\newcommand{\var}[1]{\textit{#1}}
\newcommand{\op}[1]{\textsl{#1}}

\newcommand{\tup}[1]{%
  \relax\ifmmode%
    \langle#1 \rangle%
  \else
    $\langle$#1$\rangle$%
  \fi}

\newcommand{\CO}{\ensuremath{\mathcal{O}}\xspace}

\newcommand{\CS}{\ensuremath{\mathcal{S}}\xspace}
\newcommand{\CT}{\ensuremath{\mathcal{T}}\xspace}

\setcopyright{none}

\renewcommand\footnotetextcopyrightpermission[1]{}
\begin{document}

\title{Blockchain and Trusted Computing: Problems, Pitfalls, and a Solution
  for Hyperledger Fabric}

\author{Marcus Brandenburger}
\affiliation{IBM Research - Zurich}
\author{Christian Cachin}
\affiliation{IBM Research - Zurich}
\author{R\"udiger Kapitza}
\affiliation{TU Braunschweig}
\author{Alessandro Sorniotti}
\affiliation{IBM Research - Zurich}

\begin{abstract}
  A smart contract on a blockchain cannot keep a secret because its data is
  replicated on all nodes in a network.  To remedy this problem, it has
  been suggested to combine blockchains with \emph{trusted execution
    environments} (TEEs), such as Intel SGX, for executing applications
  that demand privacy.  As a consequence untrusted blockchain nodes 
  cannot get access to the data and computations inside the TEE.

  This paper first explores some pitfalls that arise from the combination
  of TEEs with blockchains.  Since smart contracts executed inside 
  TEEs are, in principle, stateless they
  are susceptible to rollback attacks, which should be prevented to
  maintain privacy for the application.  However, in blockchains with
  non-final consensus protocols, such as the proof-of-work in Ethereum and
  others, the contract execution must handle rollbacks \emph{by design}.
  This implies that TEEs for securing blockchain execution cannot be
  directly used for such blockchains; this approach works only when the
  consensus decisions are \emph{final}.

  Second, this work introduces an architecture and a prototype for
  smart-contract execution within Intel SGX technology for Hyperledger
  Fabric, a prominent enterprise blockchain platform.  Our system resolves
  difficulties posed by the execute-order-validate architecture of Fabric,
  prevents rollback attacks on TEE-based execution as far as possible, and
  minimizes the trusted computing base.  For increasing security, our
  design encapsulates each application on the blockchain within its own
  enclave that shields it from the host system.  An evaluation shows that
  the overhead moving execution into SGX is within 10\%--20\% for a
  sealed-bid auction application.
\end{abstract}

\maketitle

\section{Introduction}\label{sec:intro}

Blockchain and distributed ledger technology (DLT) have received a lot of
attention recently as means to distribute trust over many nodes in a
network.  The transparency and resilience gained from the decentralized
protocol execution ensure the integrity of blockchain applications, or
\emph{smart contracts}, realized on such a ``world computer.''  However,
the proliferation of data on a blockchain directly contradicts the goal to
keep the application state confidential and to maintain privacy for its
users, a condition that exists for many intended applications of blockchain
technology.

Although cryptographic protocols (such as secure multiparty computation and
zero-knowledge proofs) offer attractive solutions for privacy on a
blockchain, they are not yet mature enough to run general-purpose
computations easily and to be widely deployed.  As a promising alternative,
the use of \emph{trusted execution environments (TEEs)} for running
blockchain applications has been proposed, especially by the industry
working on \emph{consortium blockchains}, where the consensus process is
governed by controlled nodes~\cite{coco-whitepaper,corda-whitepaper}. 
Intel's \emph{Software Guard Extensions (SGX)} is the most prominent TEE
technology today and available together with commodity
CPUs~\cite{anati2013,hoekstra2013usinginnovative,mckeen2013}.  It
establishes trusted execution contexts called \emph{enclaves} on a CPU,
which isolate data and programs from the host operating system in hardware
and ensure that outputs are correct.  An enclave might run only a small
dedicated part of an 
application~\cite{hoekstra2013usinginnovative,mw16seckeeper,lind2017glamdring} 
or can contain an entire legacy system,
including some operating-system 
support~\cite{baumann2014,scone-osdi16,graphene-sgx}.
By using a TEE, one does not have to trust the host system of the enclave,
which runs the blockchain node and participates in protocols.

For protecting smart contracts through TEEs, issues around rollback
attacks, state continuity, and protocol integration for
TEEs~\cite{strackx2014} must be addressed.  As is well-known, there are
non-trivial interactions between integrity violations and information
leakage for stateful secure computation~\cite{AlgesheimerCCK01}.

\paragraph{Motivating example}
Imagine an auction of a digital item on a blockchain.  In sealed-bid
auction designs~\cite{milgrom04} (e.g., Vickrey auctions) keeping the bids
secret is of primary importance, so that neither another bidder nor any
other party can learn anything about them.  Only a trusted auctioneer
should learn the bids to the extent necessary for evaluating the auction.
For moving the auction to a blockchain, the functions of the auctioneer are
implemented by a smart contract.  The distributed ledger stores encrypted
bids such that the bidders are able to verify that their submitted bids
were actually considered in the final evaluation.  The blockchain nodes
execute the auction's smart contract, which records the bids, closes the
auction, evaluates it, and autonomously executes the transaction assigning
the item to the winning bidder and transferring the payment to the seller.

By running the auction's code within an SGX enclave, the auction maintains
privacy and simultaneously benefits from the transparency of the
blockchain.  More precisely, the bids are encrypted, the key to decrypt
them resides only inside the enclave, and the smart contract controls
operations with the key.  The bidders commit their encrypted bids to the
blockchain and the enclave decrypts them for determining the winner.
However, this simplistic auction solution may leak information, as described
next, whenever a malicious node can manipulate the operation invocation order.

\paragraph{State continuity and rollback attacks}

As the industry is slowly realizing, rollback attacks on stateful
applications running in TEEs pose serious risks, unless the \emph{state
  continuity} of an application is
ensured~\cite{parno2011,strackx2014,strpie16,rote17,bclk17}.  For instance, if a
malicious blockchain node may influence the order in which transactions
are executed by the enclave, the node can break the confidentiality of the
sealed-bid auction even if it cannot decrypt the bids.  In particular, the
node might cause the enclave to execute the evaluation transaction multiple
times and reset the enclave again afterwards, every time when a new
bid has been stored on the blockchain.  Thereby the node could learn
information about other bids.  This illustrates how an integrity violation
can lead to breaking confidentiality.
(Although platforms like SGX provide access to non-volatile monotonic
counters that might prevent rollback attacks, their use introduces
considerable complications for tolerating crashes and they are often too
slow~\cite{parno2011,bclk17}.  Hence, we do not consider them in this work.)

Rollback attacks can be prevented if the state input to the
smart-contract enclave always corresponds to the unique, committed
blockchain state.  One way to guarantee the desired state continuity would
be to run the whole blockchain node, especially its protocol logic and the
state maintenance, within the enclave.  This is often not feasible for
practical reasons, however, and leads to other security issues because the
code running inside the TEE has a large attack surface.
However, in blockchain systems with \emph{non-final} consensus protocols
that may fork temporarily, a node remains prone to being rolled back
\emph{by design}, even when it resides completely in a TEE, because the
underlying consensus protocol requires it.

\paragraph{Contributions}

In this paper we examine the state-continuity problem for trusted execution
on blockchains, arising from rollback attacks that malicious nodes might
mount.  We discuss why blockchains with consensus that has no final
decisions, such as the ``proof of work'' in Bitcoin or Ethereum, are
inherently unable to benefit from TEEs to maintain confidentiality.  If the
blockchain nodes hosting TEEs can access the final blockchain state in a
trusted way, on the other hand, then such rollbacks can be prevented.

As the main contribution of this work, we design a secure solution for
secure smart-contract execution on a blockchain using Intel
SGX, the most
prominent TEE technology available today, and \emph{Hyperledger
  Fabric}~\cite{fabric18}, or \emph{Fabric} for short, a flexible and
modular platform for consortium blockchains.  Fabric uses a modular notion
of consensus whose outputs are always final, which avoids the
protocol-inherent rollback attack.  As Fabric is the most prominent
technology for consortium blockchains today, our design can also be
integrated with other, similar systems.

Some consortium blockchain platforms follow the generic approach to
state-machine replication~\cite{schnei90}, where a consensus protocol first
decides on an order among all transactions and the nodes
\emph{subsequently} execute them according to the decided order.  In
Fabric, however, the peers execute transactions and compute state updates
\emph{before} their relative order has been determined through a consensus
protocol.  The ordering process only uses the outcome of the transaction
(i.e., the induced state changes) during consensus. While this offers a 
flexible programming model for smart contracts~\cite{fabric18}, it 
also introduces additional complications that must be considered.

We have implemented a prototype that enables smart-contract execution inside
Intel SGX for Hyperledger Fabric.  We demonstrate an auction application
and evaluate the performance of the prototype compared to the unprotected
execution.  The results show that our prototype reaches 0.80x--0.90x the
throughput of the unprotected implementation, which is acceptable for
protecting the confidentiality.

\paragraph{Organization.}

This paper is structured as follows.  In Section~\ref{sec:pow} we discuss
why public blockchains with non-final consensus are inherently unable to
execute smart contracts in TEEs.  Section~\ref{sec:preliminaries}
introduces Intel SGX and Fabric, the two technologies used mostly in the
remainder of the paper.  In Section~\ref{sec:problem} we introduce the
system model, describe the security goals, and discuss several approaches
to run smart contracts in TEEs and their complications.  Our solution
to execute applications on Fabric with SGX is presented in
Section~\ref{sec:approach} and its security is examined in
Section~\ref{sec:security}.  Performance evaluation results are reported in
Section~\ref{sec:evaluation}.  Finally, Section~\ref{sec:related} reviews
related work and Section~\ref{sec:conclusion} concludes the paper.

\section{Consensus with non-final decisions}
\label{sec:pow}

Public blockchains patterned after Bitcoin~\cite{nakamo09} do not reach
consensus with \emph{finality}.  Their consensus mechanism is based on a
randomized protocol, in which for each epoch (or ``block height'') a node
selected through a probabilistic scheme that is difficult to bias (such as
a ``proof of work'') disseminates a block of transactions to be appended to
the blockchain.  Such blocks are propagated to all nodes with a
peer-to-peer gossip protocol that is efficient but does not guarantee
strict consistency.  During regular operation, the view of the nodes in
different parts of the network may diverge, and such forks are resolved
through the protocol rule that the ``longest'' branch is adopted by all
nodes as the valid blockchain and determines the state.  As shown by
multiple formal analyses of the protocol (e.g.,~\cite{gakile15}), the
probability that such forks last for many epochs vanishes exponentially
fast, but it cannot be made negligible for short forks.  If the underlying
network does not ensure universal connectivity, this can lead to
devastating attacks on the safety of a public blockchain~\cite{apzova17}.

When a node first receives the block as a candidate that should extend the
current chain, the node validates the block's content, including that all
transactions inside are correct.  For Bitcoin this validation is simply
checking that a ``coin state'' has not been spent earlier, but for
programmable blockchains like Ethereum~\cite{EthereumYellowPaper}, this
entails executing all transactions and computing the corresponding state
updates.  If the block is valid, the node appends the block to its local
chain and updates its state accordingly.  But when the node later receives
other blocks that are all valid and collectively extend a prefix of the
currently held chain to a ``longer'' chain, the node reverts the earlier
transactions and instead executes the subsequently received transactions.
There is a significant probability that a node has to revert transactions
during regular operation and, therefore, consensus is never final.
In essence, a node must continue to participate in the consensus protocol
forever, just to be sure that the blockchain state it holds remains valid.
Blockchains using ``proof-of-stake'' consensus also suffer from similar
forks (see the overview and analysis by David et al.~\cite{dgkr18}).

As becomes clear from this discussion, TEEs cannot be used to secure
transaction execution and validation in blockchains based on non-final
consensus.  For example, an Ethereum virtual machine (EVM) running within
SGX would have to produce the outputs resulting from a transaction
immediately, but already during normal operation, the EVM could be rolled
back to an earlier state that is beyond its control.  This also holds if
the consensus protocol is executed inside the TEE, as a malicious host
controlling the communication could censor blocks from the network and
forge valid blocks of its choice, given enough time.  As argued before
through the auction example, application-level secrets could be revealed
easily.

Therefore consensus with finality seems to be a necessary prerequisite to
rely on TEEs for securing blockchains and for keeping transaction data
secret.  If one lets the TEE execute only transactions that are final, any
attempt to roll back its state amounts to an attack, and such attacks can
be prevented using existing methods for state continuity.
This insight stands also behind some of the early designs and technologies
that aim at this goal.  For example, Microsoft's Coco
Framework~\cite{coco-whitepaper}, available only as a white paper so far,
uses the EVM but mentions quorum-based consensus with finality.  In the
Hyperledger Sawtooth platform
(\url{http://www.hyperledger.org/projects/sawtooth}), which is most actively
developed by Intel, the role of SGX technology lies in securing the
``proof-of-elapsed-time (PoET)'' consensus protocol, but SGX is not used
for safeguarding secrets of a smart contract.

\section{Technologies}
\label{sec:preliminaries}

In this section we review Hyperledger Fabric, an open-source blockchain
platform developed under the Hyperledger Project
(\url{http://www.hyperledger.org}) hosted by the Linux Foundation.
We then describe \acf{SGX}, which adds hardware-enforced security to the Intel
CPU architecture and enables secure smart contract execution.

\subsection{Hyperledger Fabric}
\label{sec:back:fabric}

Hyperledger Fabric~\cite{fabric18} is a permissioned blockchain platform (run
by a consortium), where multiple parties may participate and together form a
distributed ledger network.  The ledger records all interactions between the
parties as transactions.  A transaction invokes a smart contract called
\emph{chaincode}, which defines an application running on the blockchain.

\begin{figure}[ht!]
    \centering

    \includegraphics[width=7cm]{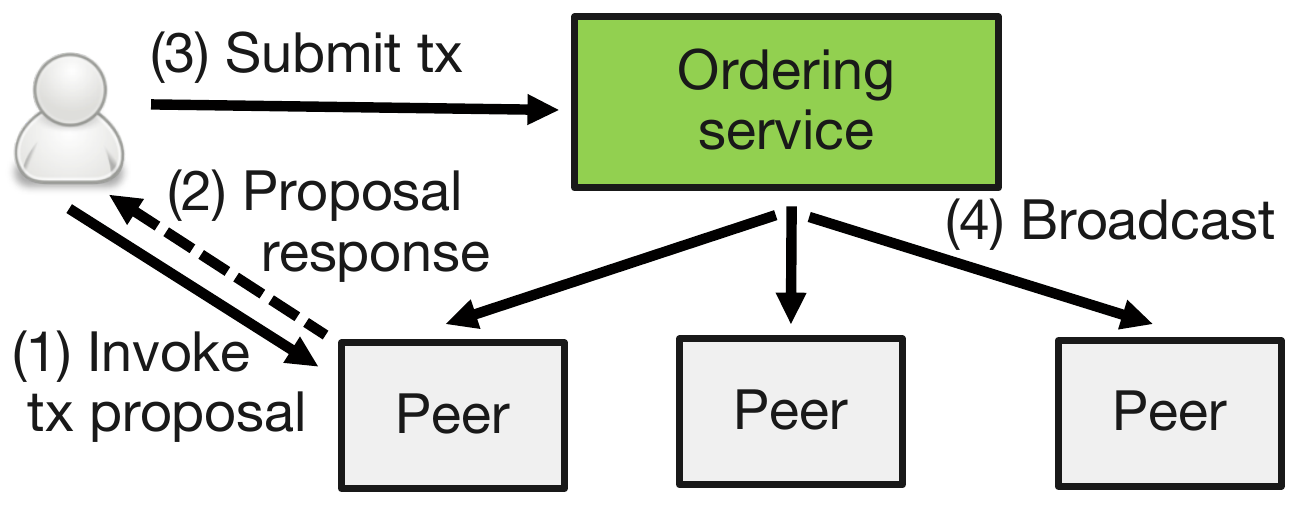}
    \vspace*{-3mm}
    \caption{A high-level view of a Fabric network, illustrating the transaction flow invoked by a client (left).}
    \label{fig:fabric_arch}
    \vspace*{-3mm}
\end{figure}

A Fabric network consists of \emph{clients}, \emph{peers}, and an
\emph{ordering service}, as illustrated in Fig.~\ref{fig:fabric_arch}.  For
each peer, a special client called \emph{admin} has administrative control
over the peer, for instance in order to install a chaincode.
The basic transaction flow is as follows:
\begin{enumerate*}
  [label=(\arabic*)]
  \item A client invokes a chaincode by sending a transaction proposal to one
  or more peers, which

  \item execute the chaincode and produce a proposal response called
  \emph{endorsement}.
 
  \item The client then collects the endorsements and assembles them
  to a transaction that it submits to the ordering service.
 
  \item The ordering service establishes the total order of all transactions
  and broadcasts them as blocks of transactions to all peers in the network.
  When a peer receives a block, it validates every transaction, eliminates
  those that were based on state that has become invalid,
  and commits the valid ones to its local ledger.
\end{enumerate*}

Other blockchain platforms execute transactions after ordering
them~\cite{cacvuk17}, e.g., JPMC Quorum
(\url{https://github.com/jpmorganchase/quorum}), Hyperledger Sawtooth
(\url{https://github.com/hyperledger/sawtooth-core}), or Chain Core
(\url{https://chain.com/}).  In contrast, Fabric uses a three-phase
\emph{execute-order-validate} architecture.  In the remainder of this
section we provide more details of each phase.

\paragraph{Chaincode execution and endorsement.}
A transaction of a chaincode is executed by a set of \emph{endorsing peers} for
the chaincode during the endorsement phase.
Initially the chaincode is installed on every endorsing peer by an admin.
The clients invoke the chaincode by sending a transaction proposal
containing a chaincode operation to the peer.  The chaincode takes the
operation as input, processes it according to its smart-contract
application, and may return an optional execution result.

While executing, the chaincode may access the blockchain state, which is
provided as a \emph{key-value store (KVS)}, with \op{getState} and
\op{putState} operations.  The \emph{putState} operation does not
immediately update the peers' local state; instead, it records the change
in a \emph{writeset}, containing the updated keys and their values.
Additionally, all keys accessed by the transaction during execution and
their versions (i.e., the positions in the history where they were last
updated) are collected in a so-called \emph{readset}.

Taken together, the execution result, the readset, and the writeset
form a transaction proposal response or \emph{endorsement}.
The peer returns this to the client.
Through executing the chaincode and producing the endorsement, the peer
vouches for the correct execution of the chaincode and \emph{endorses} the
change.
An endorsement policy specifies the endorsing rules for each
chaincode~(e.g., who are the endorsing peers or how many endorsements are
needed).  Accordingly, the client collects sufficiently many endorsements
and integrates them to a transaction that it submits for ordering.

\paragraph{Ordering.}
The ordering service in Fabric is responsible for establishing the
\emph{total order} of all transactions in the blockchain and therefore to
ensure a consistent view of all transactions across all peers.  Typically,
the ordering service consists of multiple nodes for scalability and
resilience, and leverages a protocol to reach consensus on the total order.
Clients submit transactions created in the endorsement phase to
the ordering service.  For efficiency, transactions are distributed among all
peers in batches or \emph{blocks}, using a gossip
protocol~\cite{gossip-fabric}.  

\paragraph{Validation and state updates.}
A peer receives a block of transactions from the ordering service and
utilizes a \emph{validation system chaincode (VSCC)} to validate each
transaction and apply its effects to the local ledger. 
Validation is a deterministic process and performed by every peer. In
particular, the peer checks that every transaction fulfills the endorsement
policy.
Then, a \emph{read-write conflict} check is performed, that is, the peer
verifies that the versions in the readset match the current blockchain state
at the peer.
If both validation checks are successful, the updates in the writeset
of the transaction are applied to local blockchain state of the peer.
An invalid transaction has no effect on the state and
the issuing client should reinvoke the transaction again.
A Fabric blockchain is initialized through a so-called \emph{genesis block},
which is created collaboratively by all network participants.

\subsection{Trusted execution with Intel SGX}
\label{sec:back:sgx}

Modern \acfp{TEE}, such as
\acf{SGX}~\cite{anati2013,hoekstra2013usinginnovative,mckeen2013}, add
hardware-enforced security to commodity platforms.  \ac{SGX} enables Fabric
peers to execute chaincode in a trusted execution context, also called
\emph{enclave}.
Particularly, an enclave defines an isolated memory area that is guarded by
hardware-enforced mechanisms, which guarantee \emph{confidentiality} and
\emph{integrity} of an enclave even if the entire platform is compromised.
That is, even higher-privileged code (e.g., the operating system) can neither
access that memory area nor modify it without being detected.

\paragraph{Enclave protection and attestation.}

\ac{SGX} enforces that only genuine applications are executed in an enclave.
``Genuine'' means that the code has not been tampered with and operates
precisely as intended by the developer.
For this reason, a cryptographic hash called $\var{mrenclave}$ of the
code and data initially loaded into the enclave is generated by the CPU.  If $\var{mrenclave}$
matches the hash of the genuine application signed by the 
developer, the enclave starts successfully.
This ensures that the correct application (e.g., a specific chaincode) is
runs in an enclave.

During runtime, an enclave is capable to prove to a third party (e.g., a
Fabric client or another peer) that a specific application is loaded and
executed in an actual enclave on a \ac{SGX} platform.  For this purpose a
procedure called \emph{remote attestation}~\cite{anati2013} is used.
It works as follows: A client with prior information about
$\var{mrenclave}$ of the target enclave sends an attestation challenge to the
enclave host (e.g., the peer) and in return receives a proof~$\phi$, also
known as \emph{quote}, produced by a target enclave and the platform.
We use the term \emph{attestation report} to refer to a quote.
The client forwards~$\phi$ to the \ac{IAS}, which verifies it using a
\emph{group signature} scheme called EPID~\cite{brickell2011enhanced},
and replies with an attestation result.
The attestation result either confirms that~$\phi$ was actually produced by
an \ac{SGX} enclave running the intended code or indicates that~$\phi$ is
invalid.  The enclave can also embed custom data in~$\phi$, which builds
the basis for key exchange protocols.

Remote attestation involves an intermediate step called \emph{local
attestation} that is performed between two enclaves on the same platform.
In this step, a special enclave called \emph{Quoting Enclave~(QE)} verifies
that~$\phi$ was produced by an enclave on the same platform, 
using HMAC with a shared secret key only accessible by enclaves on
the same platform.  Local attestation can be performed by any two enclaves on
the same platform.

\paragraph{Enclave state and data sealing.} 
Since enclaves reside in a protected memory area in the CPU, enclaves are
volatile, thus, when an enclave stops, restarts, or just crashes, its internal
state is lost and cannot be recovered.  For this reason \ac{SGX} supports
data sealing, a mechanism that allows to encrypt and authenticate data before
it leaves an enclave and is stored externally (e.g., on persistent storage).
After an enclave restarts, it may load the sealed data and decrypt it.
Although the sealing mechanism protects data confidentiality, it does not
prevent rollback attacks, that is, an attacker may cause an enclave to recover from properly
sealed but stale data~\cite{bclk17,rote17,strpie16}.
For many applications this poses serious problems and they must 
be protect against it, however, for stateless applications, such as
chaincodes, this is not relevant.

Moreover, enclaves have access to a secure random number generator
that allows to build cryptographic primitives, such as key generation,
encryption, and digital signatures.  Both, remote attestation and data
sealing rely on a cryptographic key-management infrastructure rooted in a
secret key fused into the CPU, which provides deterministic key-derivation
functions to an enclave.

\section{Problem description}
\label{sec:problem}

This section describes the problem of secure smart-contract execution using
trusted hardware for blockchains with \emph{final} consensus, in
particular, Hyperledger Fabric~\cite{fabric18}.  We explore intricacies
that may still be caused by rollback attacks in this setting, illustrate a
strawman approach that is infeasible, and introduce our approach to support
secure chaincode execution using Intel \ac{SGX}.  This executes each
chaincode in its own enclave during endorsement at a peer and thereby
protects the confidentiality and integrity of the blockchain application.

\subsection{System model}
\label{sec:model}

We consider a Fabric blockchain network with clients, an ordering service,
and a set of peers, which collaboratively execute transactions and maintain
a distributed ledger on a single Fabric ``channel.''

A client  invokes transactions by sending a chaincode operation to some
peer, which then executes (simulates) it and produces an endorsement
containing the resulting state change on the ledger.  The operation, the
response, as well as the ledger may contain sensitive information that
should stay secret.

To prevent such information leakage, every peer is equipped with an
\ac{SGX}-enabled CPU and executes transactions inside an enclave.
The chaincode is stateless, and a transaction only takes the operation and
the blockchain state in the KVS as inputs, accessed with \op{getState}.
The chaincode must perform updates to the ledger only through \op{putState}
operations.  The execution of a chaincode operation returns a response that
may include a computation result, the state update, and the read-write
dependencies.

\subsection{Threats}
\label{sec:threats}

Although most peers are usually correct, a peer may become \emph{malicious}
and behave incorrectly, for instance, when it tries to maximize its own
profit or becomes corrupted by an attacker.
A peer has full control over the operating system, applications, and the data
residing in memory and persistent storage~(i.e., the blockchain state).
A malicious peer, however, cannot access or tamper with the code and data
residing in an enclave (see Section~\ref{sec:back:sgx}). A malicious peer
may neither break cryptographic primitives nor extract any secret
information that from an enclave.
Consequently, a chaincode running in an enclave always produces the correct
results, that is, the chaincode does not deviate from its specification,
the enclave-internal state is only known to the enclave itself, and nothing
is revealed apart from the resulting state change.

However, a malicious peer can invoke the chaincode enclave with any input
and in arbitrary order.
The peer may intercept, modify, reorder, discard, or replay chaincode
operations, and when the chaincode enclave accesses the KVS, the peer may
feed any blockchain state to it.

Furthermore, the peer might even drop messages or completely halt an
enclave, but we do not consider such denial-of-service attacks in this
work.
We also ignore potential information leakage from SGX on side
channels~\cite{BrasserMDKCS17,schwarz2017malware,shih2017t,xu2015controlled}
because this appears orthogonal to our focus.

As is well-known from the literature on secure computation with
cryptographic protocols~\cite{AlgesheimerCCK01}, integrity and
confidentiality cannot be considered separately.  Likewise, for a secure
application running in an enclave, a malicious host may break
confidentiality by triggering the enclave to execute on ``incorrect''
inputs.  In the blockchain context, this means that the chaincode
execution deviates from the consensus-based transaction order. 

Repeating and extending the auction example from the introduction, such an
attack could reveal secret information as follows.  Suppose that evaluating
the auction on the current blockchain state~$s_1$ would let a bid~$b_1$ win
the auction.  If the malicious node can trigger the auction-evaluation
transaction, it learns $b_1$.  If the node can reset the enclave to $s_1$
and execute another transaction, it can submit a bid~$b_2$, add it to the
ledger, subsequently evaluate the auction, and learn if $b_2 > b_1$.
Such a rollback attack clearly breaks the confidentiality of the individual
bids.  As mentioned earlier, rollback attacks on trusted execution
environments and their prevention has only recently been understood
better~\cite{strpie16,bclk17,rote17}.

\subsection{Strawman approach}
\label{sec:strawman}
It follows from the discussion in Section~\ref{sec:pow} that letting the
enclave only execute transactions that have been ordered by the network
with finality prevents the rollback problem.  This amounts to running the
entire blockchain peer inside an enclave, as also suggested by Microsoft
Coco~\cite{coco-whitepaper,msr-volt} and related work.
We call this the \emph{strawman approach} that might work for an order-execute
architecture where the consensus process has only final decisions, but we
argue later why better designs exist.

For Fabric, the strawman design would mean to encapsulate the chaincode
execution, endorser, committer, ledger-state access, and all other parts of
a peer inside an enclave.  This obviously protects the integrity of the
input sequence for the chaincode, since the entire Fabric peer runs within
\ac{SGX}.
A similar approach is taken in the blockchain-as-a-service platform of IBM,
which deploys Fabric peer as a \emph{secure service container} on an IBM Z
system.  The secure container includes the whole operating system,
middleware stack, and blockchain platform~\cite{ibm-ssc}.

Although no operating system is running within SGX, recent research has
demonstrated how legacy applications can run in \ac{SGX} through a
library OS that executes unmodified applications in an
enclave~\cite{baumann2014,scone-osdi16,graphene-sgx}.  Note that the
library OS adds tens of thousands of lines of code that also run along
the application in the enclave.

This approach introduces multiple problems, however.
First, it stands in contrast to the important computer-security principle
of minimizing the size of the \emph{trusted computing base (TCB)}.
Specifically, also the \ac{SGX} developer guidelines~\cite{sgx-guide}
recommend to partition an application into a trusted and an untrusted
component; only a small portion of the application code should execute
inside the enclave.  A smaller TCB has fewer errors, reduces the attack
surface, and is more amenable to security analysis than the entire
application.

A second problem stems from the limited memory available to enclaves.  An
enclave's memory resides in the \emph{enclave page cache (EPC)} isolated
from the rest of the system.  The EPC is currently limited to 128 MB.  Once
an enclave reaches that limit pages are outsourced to DRAM.
This results in a dramatic loss of performance, as reported in several
works~\cite{scone-osdi16,mw16seckeeper,orenbach2017eleos}.
In particular, since the ledger grows with every block, holding the whole
blockchain state in the enclave quickly reaches the memory limitation.

\vspace*{-1mm}
\subsection{Approach for Hyperledger Fabric}

To avoid the drawbacks of the strawman approach, we adopt a modular
architecture that separates the chaincode execution conceptually from the
peer and moves the execution into an enclave.  The
protocol-specific aspects of the peer are encapsulated in an abstract
ordering service, of which one process might run on the same peer.
The ordering service is \emph{trusted} in the sense that it cannot be
rolled back.

The ordering service produces a signed sequence
of transactions for execution within the enclave.
The enclave can verify that transactions originate
from the ordering service, are in the proper order, and have not been
tampered with.
The enclave also keeps information about the transaction history, which
allows to detect transaction-ordering violations or replayed transactions.
The malicious host might still reset the enclave to an earlier point in the
execution sequence, but this would not harm the application since the
transactions are deterministic and execution would simply produce the same
outputs again.

As described so far, this approach works fine with an order-execute
architecture for state-machine replication.  Fabric, however, uses the
execute-order-validate paradigm, where a peer executes a transaction before
consensus on the order is reached (see Sec.~\ref{sec:back:fabric}).
Consequently the execution is speculative and can be repeated without
affecting the blockchain state, as transactions are simulated during
endorsement and only take effect after the ordering.
This means a malicious host could infer information about the secret
application data from the speculative execution.  Not even a trusted
ordering service can prevent this type of leakage.

To resolve this issue, we will have to adapt the applications to respect
the speculative nature of execution in Fabric.  For the auction example, in
particular, a \emph{barrier} will be stored on the blockchain such that the
chaincode enclave only evaluates the auction if the barrier is present.
The barrier is set by invoking the chaincode with a transaction to ``close''
the auction but not yet evaluate it.
If the barrier is present on the ledger, a malicious peer may no longer
submit new bids to the auction.  On the other hand, the auction evaluation
will only consider bids added to the ledger before the barrier.  Note that
this barrier plays a role similar to a memory barrier in a multi-core
computer system with concurrent threads.

Following the execute-order-validate architecture, the chaincode enclave
must execute transactions only on the committed blockchain state, that is,
with ledger entries that result from ordered transactions and that have
been committed by all peers.
Otherwise a malicious peer may produce the barrier itself and feed the
resulting state into the enclave when evaluating the auction.
The system described in the next section ensures this.

To formally model the information leakage permitted in the
execute-order-validate architecture of Fabric, we model a blockchain as
stateful functionality~$F : \CS \times \CT \to \CS$.  At any time the state
of the chaincode is an element of \CS.  The clients invoke transactions in
$\CT$, which may contain operations with arguments according to~$F$, but
these are subsumed into the different~$t \in \CT$.
Given~$s \in \CS$, applying a transaction $t \in \CT$ of~$F$ means to
compute~$s' \gets F(s,t)$, resulting in a subsequent state~$s' \in \CS$.
Using a trusted ordering service as introduced earlier, the blockchain's
state evolution is defined through the sequence of transactions signed by
ordering.

With the chaincode functionality~$F$ running in an SGX enclave, even a
malicious peer may only learn the subsequent state resulting from a
transaction, but nothing about the computation itself.  Since cryptographic
keys could reside in the enclave, the ledger state doesn't necessarily
reveal all relevant information.  Due to the rollback attacks introduced
earlier, however, such a peer can execute any transaction \emph{on any
  input state} that is in the history of transactions issued by the
ordering service.

\begin{definition}[Security up to resets]
  Consider a blockchain system with an execute-order-validate architecture
  and suppose the correct ordering service produces a sequence of states
  $\langle s_0, s_1, \dots,s_m \rangle$, where $s_{j} = F(s_{j-1}, t_j)$
  for $t_j \in \CT$ and $j \in [1,m]$.  We say that the chaincode is
  \emph{secure up to resets} if any malicious peer, through interacting
  with the chaincode running inside the enclave, may obtain states
  $s^*_{k+1} = F(s_{k},t^*)$, for any $k \in \{0, 1, \dots, m\}$ and an
  arbitrary transaction~$t^* \in \CT$, but no further information.
\end{definition}

The security-up-to-resets notion formalizes attacks on TEE-based execution
in Fabric, where a malicious peer may collude with a client.  The client
invokes an arbitrary transaction~$t^*$ that reveals information about the
chaincode's state.  The peer lets the TEE execute $t^*$ and produce an
output, but the endorsement is never sent for ordering and the transaction
is never appended to the blockchain.  The chaincode may leak all states
resulting from such executions.

Note that Fabric permits parallelism during execution for separating trust
assumptions and for increasing scalability.  By adding a barrier into the
blockchain, an application essentially benefits from the guarantees of the
order-execute design with respect to rollbacks across the barrier.
Requiring a barrier after every transaction would actually impose the
order-execute paradigm onto Fabric.

\section{Secure chaincode execution}
\label{sec:approach}
This section describes our system for secure chaincode execution in
Hyperledger Fabric using Intel \ac{SGX}.

\subsection{System architecture}
\label{sec:arch}
\begin{figure}[ht!]
    \centering
   \includegraphics[width=7.4cm]{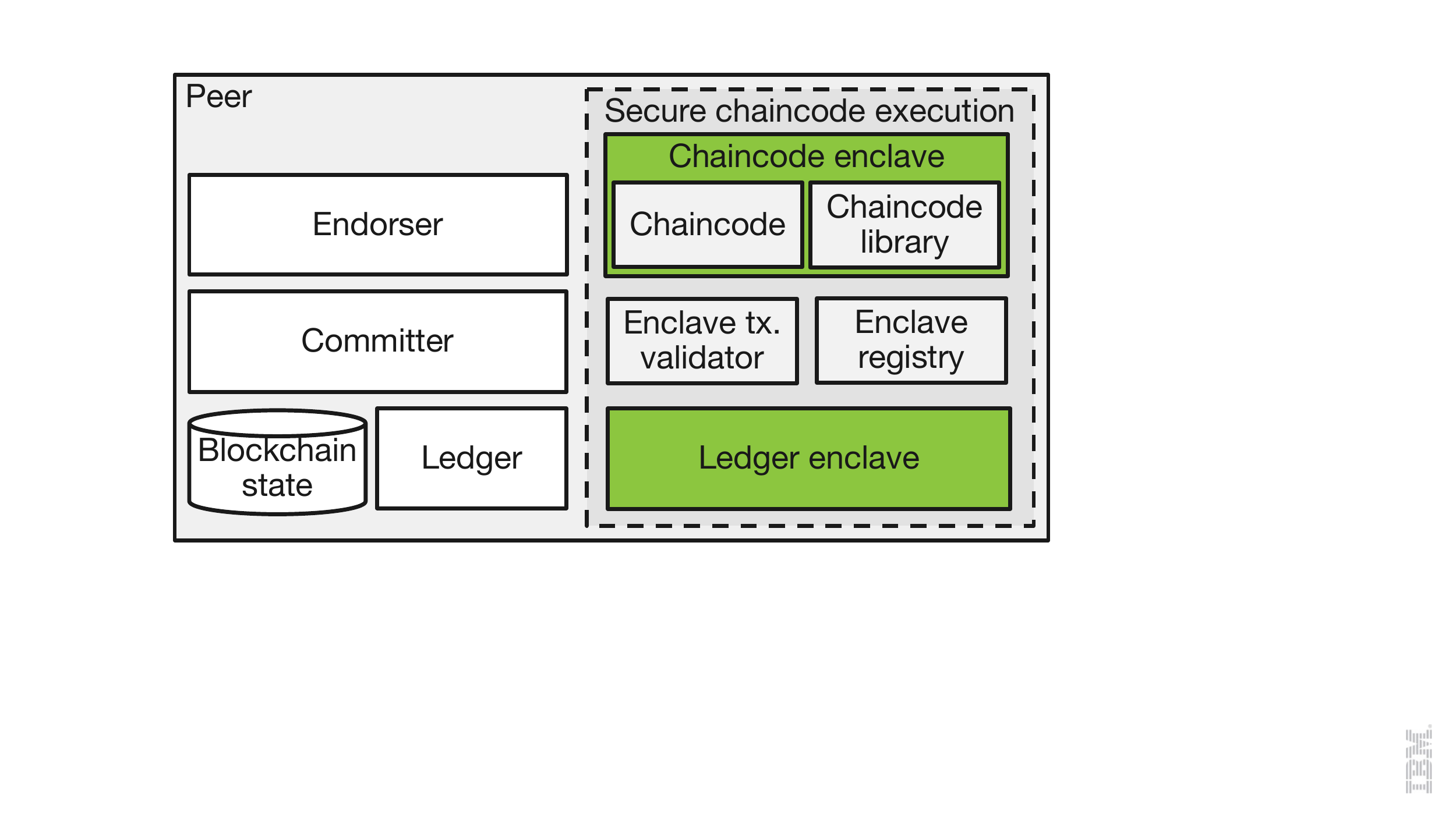}
    \vspace*{-3mm}
    \caption{System architecture. The dashed box denotes the components
      added to the peer to enable secure chaincode execution with
      \ac{SGX}. The components running within \ac{SGX} enclaves are denoted
      in green (or dark) color.} 
    \label{fig:comp:arch}
    \vspace*{-3mm}
\end{figure}

Our approach extends a Fabric peer with the following components:
A \emph{chaincode enclave} that executes a particular chaincode and a \emph{ledger
  enclave} that enables all chaincode enclaves to verify the blockchain
state integrity; all run inside \ac{SGX}.  In the untrusted part of the
peer, an \emph{enclave registry} maintains the identities of all chaincode
enclaves and an \emph{enclave transaction validator} that is responsible for validating
transactions executed by a chaincode enclave before committing them to the ledger.
Fig.~\ref{fig:comp:arch} shows the components.

\paragraph{Chaincode enclave.}
The chaincode enclave executes one particular \emph{chaincode}, and
thereby isolates it from the peer and from other chaincodes.
A \emph{chaincode library} acts as intermediary between the chaincode in
the enclave and the peer.  The chaincode library exposes the Fabric
chaincode interface and extends it with additional support for state
encryption, attestation, and secure blockchain state access.

\paragraph{Ledger enclave.}
The ledger enclave maintains the ledger in an enclave in the form of
integrity-specific metadata representing the most recent blockchain state.
It performs the same validation steps as the peer (see
Sec.~\ref{sec:back:fabric}) when a new block arrives, but additionally
generates a cryptographic hash of each key-value pair of the blockchain state
and stores it within the enclave.
The ledger enclave exposes an interface to the chaincode enclave for
accessing the integrity-specific metadata.  This is used to verify the
correctness of the data retrieved from the blockchain state.

\paragraph{Enclave registry.}
The enclave registry is a chaincode that runs outside \ac{SGX} and maintains a
list of all existing chaincode enclaves in the network.  It performs
attestation (see Sec.~\ref{sec:back:sgx}) with the chaincode enclave and
stores the attestation result on the blockchain.  The attestation demonstrates
that a specific chaincode executes in an actual enclave.  This enables the
peers and the clients to inspect the attestation of a chaincode enclave before
invoking chaincode operations or committing state changes.

\paragraph{Enclave transaction validator.}
The enclave transaction validator complements the peer's validation system and
is responsible for validating transactions produced by a chaincode enclave.
In particular, the enclave transaction validator checks that a transaction
contains a valid signature issued by a registered chaincode enclave.  If the
validation is successful, it marks the transactions as valid and 
hands it over to the ledger enclave, which crosschecks the decision  
before it finally commits the transaction to the ledger.

\subsection{System initialization}

When a peer joins the blockchain network, the ledger enclave is initialized
by the admin with the genesis block, which contains the blockchain
configuration and the expected hash ($\var{mrenclave}$) of the ledger enclave.
If the actual $\var{mrenclave}$ obtained by the peer does not match the value
in the genesis block, the ledger enclave does not proceed with the
initialization.
The ledger enclave then generates a private/public key
pair~$(\var{SK}_{LE},\var{PK}_{LE})$, which allows to uniquely identify the
ledger enclave.  The public key is revealed to the chaincode enclaves whereas
the private key is kept secret within the ledger enclave.
 
The ledger enclave maintains several configuration values initially obtained
from the genesis block, such as the identities (i.e., public keys) of the
peers, the clients, and the ordering service, which are used to authenticate
all received blocks and transactions.
The ledger enclave only accepts blocks that come from the ordering service as
defined in the genesis block.  To ensure this, it verifies that each block
has a valid signature issued by the ordering service.
Note that the blockchain consortium configuration can be updated using
configuration transactions.  For simplicity, however, we assume a static
consortium.

Every block has a sequence number and contains a list of transactions.  The
ledger enclave maintains information about the most recently processed
transaction, to ensure that all blocks are processed in the correct order
and no blocks are missing.

Once the peer has joined the network and has started its ledger enclave, the
peer admin also installs the enclave registry on every peer and instantiates
it. This completes the initialization of the peer.

\subsection{Chaincode enclave bootstrapping}
\label{sec:approach:bootstrapping}

We now describe how to initialize a chaincode enclave.  This is
initiated by the peer admin and consists of the following phases: (1) creating the
chaincode enclave; (2) registering with the enclave registry; (3) provisioning
of secrets; and (4) binding the chaincode enclave to the ledger enclave. These
phases are explained next.

\begin{figure}[ht!]
    \centering
    \includegraphics[width=6.4cm]{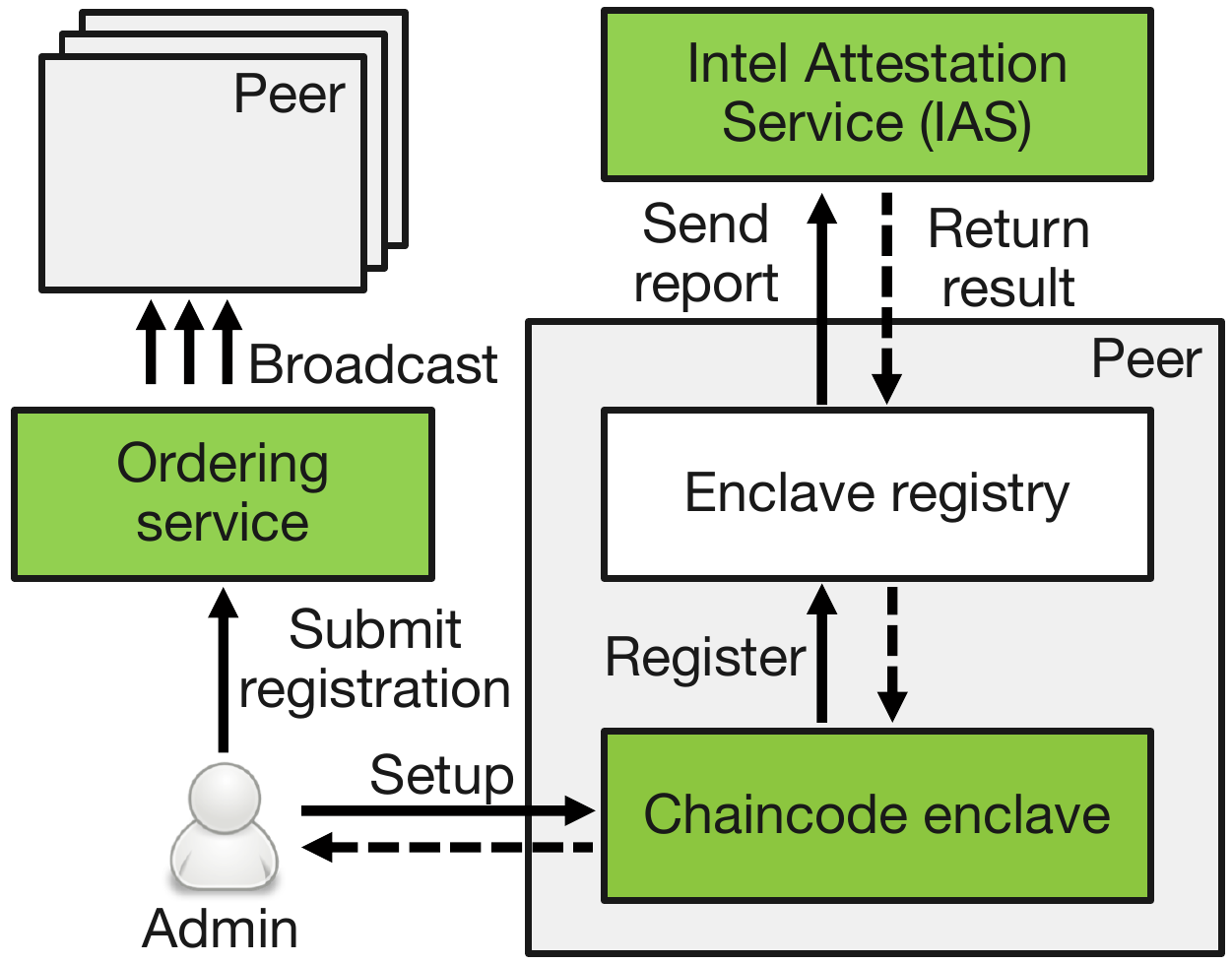}
    \vspace*{-3mm}
    \caption{Enclave registration process.}
    \label{fig:comp:ercc_vscc}
    \vspace*{-3mm}
\end{figure}

In the first phase, the admin installs the chaincode enclave at the peer and
then sends a \str{setup} transaction proposal.  The peer then starts the
chaincode enclave, which generates a private/public key
pair~$(\var{SK}_{CC},\var{PK}_{CC})$.  As for the ledger enclave, the public
key is used to uniquely identify the chaincode enclave.

Second, the chaincode enclave registers itself with the enclave registry as
shown in Fig.~\ref{fig:comp:ercc_vscc}. For this purpose, the chaincode
enclave calls $\str{register}$ and in turn, the enclave registry performs
remote attestation of the chaincode, as described in
Section~\ref{sec:back:sgx}.
In detail, the chaincode enclave first produces an \emph{attestation report}
that manifests that it is properly instantiated with a specific chaincode and
it is identified by~$\var{PK}_{CC}$.  The report contains
$\var{mrenclave}_{CC}$ and (hash of) $\var{PK}_{CC}$.  The
chaincode enclave then calls \str{register} at the enclave registry with the
report and its public key as arguments.
The enclave registry first checks that the report contains the expected
$\var{mrenclave}_{CC}$ and the correct hash of $\var{PK}_{CC}$.
Subsequently, it sends the report to the \ac{IAS} for verification and in
return receives an \emph{attestation result}, which shows whether the report
was valid or not.  Note that the attestation result is signed by the \ac{IAS}
and its verification key is publicly available.
If the verification succeeds, the enclave registry completes the registration
by calling $\op{putState}$ to store the attestation result together with
$\var{PK}_{CC}$ on the ledger.  
This makes the attestation result accessible to all peers in the network through the
ledger, certifying that this enclave runs the particular chaincode on the
given peer.  Clients and other peers use this in two ways.  First, a client
verifies that it invokes transactions involving secret data on an enclave
authorized for this.  Second, the enclave transaction validator of a peer,
which updates the blockchain state, verifies that the execution results are
genuine and result from the secure execution in the enclave.

After successfully registering the chaincode enclave, the admin optionally
provisions the chaincode enclave with secrets. For instance, the admin may
inject an encryption key for data stored on the blockchain into the chaincode
enclave (see also Sec.~\ref{sec:accessing_state}).

In the last phase, the chaincode enclave binds to the ledger enclave
through local attestation (see Sec.~\ref{sec:back:sgx}).  This means
that the chaincode enclave requests the ledger enclave to prove that it is
runs the expected ledger-enclave code and runs on the same host platform.
The ledger enclave produces an attestation report and returns it to the
chaincode enclave, which then performs the same verification steps as
described above.  (In contrast to remote attestation, the cryptographic
protection of local attestation uses HMAC and a shared key for
verification, provided by the SGX platform.)
If the verification succeeds, the chaincode enclave stores the ledger
enclave's public key~$\var{PK}_{LE}$ and thereby binds itself to the ledger
enclave, in the sense that the chaincode enclave uses this for verifying
accesses to the
blockchain state.  The chaincode enclave rejects any blockchain state
values not originating from this ledger enclave.

\subsection{Chaincode execution}

\paragraph{Endorsement.}

A client triggers the chaincode execution by sending an \str{invoke}
transaction proposal with a chaincode operation to the peer.  The peer
forwards the chaincode operation to the chaincode enclave, which then
processes it according to the smart contract.  The chaincode enclave
prepares a response and returns it to the peer, which subsequently sends it as
a a transaction proposal response to the client.

In more detail, prior to invoking the chaincode enclave, the client queries the
peer to retrieve the enclave's public key~$\var{PK}_{CC}$ and the
corresponding attestation result from the enclave registry.  The client
then verifies the authenticity of the attestation result, using the
\ac{IAS} verification key, and checks that the attestation contains the
expected~$\var{mrenclave}$ of the chaincode enclave,
matching~$\var{PK}_{CC}$.  If the verification succeeds, the client invokes
the chaincode enclave by preparing a transaction proposal for the target
chaincode. In particular, the client encrypts the
chaincode operation using~$\var{PK}_{CC}$, and then sends the proposal to
the peer, which extracts the chaincode operation and relays it to the
chaincode enclave.  Inside the enclave, the chaincode library decrypts the
operation using $\var{SK}_{CC}$ and invokes the chaincode with the
operation as argument.

The chaincode processes the operation,
produces a result, and returns it to the chaincode library.  The chaincode may
access the blockchain state using the chaincode library, which performs
verifies the accesses as described in
Section~\ref{sec:accessing_state}.

To complete the chaincode invocation, the enclave library creates a response,
signs it using $\var{SK}_{CC}$, and returns it to the peer.  The response
includes the operation, the readset and the writeset, and the execution result.
Optionally, the chaincode library encrypts the execution result before it
leaves the enclave using an encryption key provided by the client.  
The peer then sends the transaction proposal response back to the client,
which outputs the execution result, and submits the transaction to
the ordering service.

\paragraph{Validation and state update.}

The ordering service accepts transactions submitted by the clients, assigns
them to a block, and broadcasts the block to all peers in the network.  In
order to finalize a transaction, every peer validates the transaction and
updates its ledger copy.

For validating transactions produced by a chaincode enclave, the enclave
transaction validator essentially performs the same steps as the validation
system chaincode (VSCC), checking for conflicts and evaluating the
endorsement policy (Sec.~\ref{sec:back:fabric}).  Additionally it verifies
that the transaction was produced by the correct chaincode enclave as
follows.  The validator accesses the enclave registry to retrieve the
attestation result and public key for the enclave indicated by the
transaction.
Then it verifies these following the same steps as described
earlier.
Subsequently, it also verifies the enclave's signature on the transaction.
If this succeeds, the enclave transaction validator marks the transaction
as valid, the peer commits the transaction to its local ledger, and updates
the blockchain state accordingly.

\subsection{Accessing the blockchain state}
\label{sec:accessing_state}

Recall that a chaincode in Fabric must only use and access state on the
blockchain.  The chaincode library together with the ledger enclave
protects this data from manipulation by the local peer.

\paragraph{State integrity and consistency.}

\begin{figure}[ht!]
    \centering

   \includegraphics[width=7.4cm]{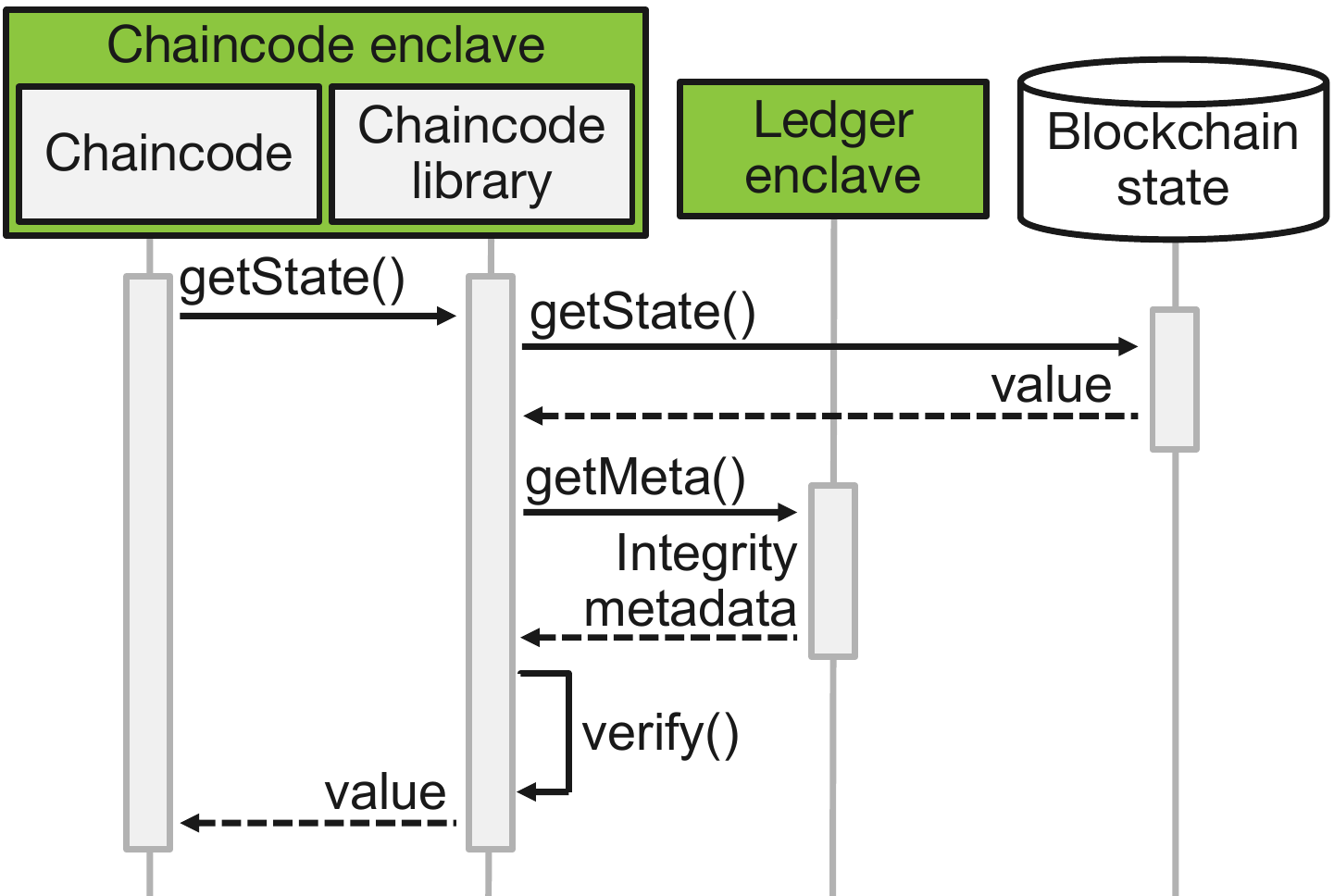}
    \vspace*{-3mm}
    \caption{Blockchain state verification with the help of the ledger enclave.}
    \label{fig:comp:tl}
    \vspace*{-3mm}
\end{figure}

As illustrated in Fig.~\ref{fig:comp:tl}, when the chaincode calls
$\op{getState}(k)$ to access the data for key~$k$, the chaincode library
loads the corresponding value~$\var{val}$ from the blockchain state
in the enclave through chaincode API provided by the peer.
Additionally, the chaincode library requests the corresponding integrity
metadata from the ledger enclave by calling $\op{getMeta}(\var{k},z)$ with
a nonce~$z$.
The ledger enclave returns the expected hash~$h_\var{val}$ of \var{val} in
the blockchain state and a signature $\phi$, produced by the ledger enclave
as $\phi = \op{sign}_{{SK}_{LE}}(k\|z\|h_\var{val})$.
The chaincode library has obtained $\var{PK}_{LE}$ during bootstrapping and
uses this to verify $\phi$.
If the signature verification over $k\|z\|\op{Hash}(\var{val})$ succeeds,
then \var{val} is correct according to the state of the ledger enclave.
The nonce ensures that the response is fresh.

\paragraph{State confidentiality.}
The chaincode library may also protect the confidentiality of the
blockchain state maintained by the chaincode enclave.
The native data sealing method of SGX for protecting persistent data
(Sec.~\ref{sec:back:sgx}) is not suitable for data shared by multiple
enclaves on different peers.  The reason is that sealed data can only be
unsealed again by the same enclave.

Instead, the chaincode library provides a state-encryption mechanism that
supports two modes: \emph{client-based encryption} and \emph{encryption per
  chaincode}.
With client-based encryption, the client is responsible for key management
and must provide an encryption key together with each chaincode operation.
For chaincode-based encryption, a chaincode-specific key must be provisioned by
an admin to all chaincode enclaves during bootstrapping.
In both modes, encryption and decryption occur transparently to the
chaincode during the $\op{putState}$ and $\op{getState}$ calls,
respectively.

As an additional benefit of client- or chaincode-based key management
compared to SGX-native methods, data on the blockchain can also be
retrieved from the blockchain later, without the support of an enclave.

\subsection{Reboot and recovery support}

A system crash or reboot terminates all enclaves instantiated on the peer.
In order to tolerate these without manual intervention, the internal states
of each chaincode enclave and the ledger enclave are stored on persistent
storage periodically.

The ledger enclave leverages sealing~(Sec.~\ref{sec:back:sgx}) to protect
its state (including integrity metadata and private key).  For ensuring
state continuity across crashes, the peer, in principle, has to write the
ledger-enclave state to disk synchronously after each block has been
processed. This clearly impacts the performance and can be mitigated
in practice by
defining a block interval for persisting the enclave state.

The chaincode enclave, in contrast, only has immutable state (including its
private key) that was created during the enclave bootstrapping.  It is
sufficient to seal and store this once, after initialization.
When the chaincode enclave restarts and restores itself from the sealed
state, it will retain the same enclave identity and does not need to
perform registration or remote attestation again.

\subsection{Extensions}

\paragraph{Support for confidential chaincode.}
Our chaincode enclave can be also extended to support the execution of
confidential chaincode, which requires support for dynamic loading 
of encrypted code in
enclaves~\cite{graphene-sgx,goltzsche17trustjs}.
This allows to deploy proprietary smart-contract code without revealing it to
the executing peers.

To enable this feature, the chaincode enclave is extended with a bootloader
that injects an encrypted chaincode binary in the enclave.  The bootloader then
decrypts it and executes the chaincode.  For this purpose, the admin that
installs the chaincode on a peer, or the chaincode developer, encrypts the
chaincode binary for a the chaincode enclave using its public key.
 
Furthermore, the attestation functionality of the chaincode enclave has to be
adapted so that the peers and clients can verify that a specific, encrypted
chaincode is executed by the enclave.  Since $\var{mrenclave}$ denotes the
bootloader code running in the enclave, the attestation must also contain a
hash of the chaincode binary, which is publicly known by the peers and
clients.

\paragraph{Trusted state transfer.}
When a peer joins an existing blockchain it has to validate all blocks processed before and
reconstruct the current blockchain state.  Depending on the age of the
blockchain, this effort may be prohibitive.  Also, a peer might have been
offline for a longer time and must catch up when it comes online again.

If the peer does not want to trust another peer for providing it the most
recent blockchain state, then it can leverage a ledger enclave to
obtain the current state securely.

When peer~$P_A$ joins or resumes after a long intermission, it contacts
another peer~$P_B$ for support.  $P_A$ sends a message containing the hash
of the genesis block and integrity metadata dependent on the position of
its ledger enclave~$LE_A$ on the blockchain.
Peer~$P_B$ passes this to its ledger enclave~$LE_B$, which performs four
steps: (1) It checks that $LE_A$ is part of the same blockchain by
comparing the hashes of the genesis blocks; (2) it calculates the
difference $\Delta$ (in terms of KVS keys) between $P_A$'s state and its
own state from $P_A$'s integrity metadata; (3) it creates an attestation
report containing $\Delta$ and its last known block sequence number; and
(4) it returns the report and $\Delta$ to~$P_B$.  At this point $\Delta$
only contains KVS keys and the corresponding integrity metadata, thus,
$P_B$ complements~$\Delta$ with the actual values from the blockchain
state.  $P_B$ also sends the attestation report to the \ac{IAS} for
verification.
Next, $P_B$ sends the attestation result and $\Delta$ to
$P_A$, and $P_A$ verifies its contents.
If successful, then $P_A$ updates its last known block sequence number and
its blockchain state accordingly, and passes the data to $LE_A$, which
performs the same verification steps as the peer and also updates the
integrity metadata.

\section{Security}
\label{sec:security}

This section argues that the secure chaincode execution system presented in
the previous section preserves security up to resets.
Recall that this security notion is defined with reference to a sequence of
blockchain states produced by transactions as decided by the trusted
ordering service~\CO.  For being secure up to resets, any set of malicious
peers interacting with SGX TEEs that host a chaincode $CC$ deployed on
Fabric must not be able to infer more than what is given from any
transaction of $CC$ invoked on one of these states.
Our informal argument proceeds in three steps.

\emph{1.\ Any state update (in the form of a writeset) produced by a
  chaincode enclave with a public key~$\var{PK}_{CC^*}$ and accepted by a
  ledger enclave into the state of its peer originates from an enclave
  whose attestation report is stored on the ledger with public
  key~$\var{PK}_{CC^*}$.
  Furthermore, any transaction output produced by a chaincode enclave, for
  which a correct client has successfully attested the output to an enclave
  with key~$\var{PK}_{CC^*}$.}

This follows from the operations of the enclave registry and the enclave
transaction validator.
In particular, the enclave registry performs remote attestation with the
chaincode enclave and thereby creates the attestation
report that it stores in the ledger.  This convinces the clients and the
peers that the chaincode enclave has been instantiated with the chaincode
represented by the $\var{mrenclave}_{CC}$ value in the ledger.  The correct
clients and peers obtain their state in the form of the sequence of blocks
with updates from~\CO and can verify the integrity of the state updates
signed by~$\var{PK}_{CC^*}$.

\emph{2.\ On any peer, the ledger state entries obtained by chaincode~$CC$
  inside an enclave represent the blockchain state after executing a prefix
  of the sequence of valid state updates that are output by~\CO.}

Note that a malicious peer may reset the ledger enclave at will to one of
the sealed and persistently stored states that the enclave produces.  Due
to the VSCC checks and the monotonically increasing sequence of block
numbers that the ledger enclave expects from \CO, the blockchain state
represented within the ledger enclave always results from executing the
sequence of transactions determined by \CO and deemed valid by VSCC and the
endorsement policies.  When the chaincode inside the chaincode enclave
accesses state in the KVS, the mutual authentication between the ledger and
the chaincode enclave, and the blockchain state-verification mechanism in
Section~\ref{sec:accessing_state} ensure that the state entries obtained by
$CC$ are correct according to the state of the ledger enclave.  Since the
ledger enclave holds the state after executing a prefix of the transaction
sequence from \CO, the above statement follows.

\emph{3.\ Any state held by a chaincode within an enclave remains
  confidential up to what is revealed by executing transactions of the
  chaincode, invoked on a prefix of the complete sequence of valid state
  updates that are output by~\CO.}

This holds because the enclaves' execution logic and data are protected
within the TEE.  Contents of the ledger enclave are sealed before they are
written to persistent storage, hence they cannot be altered by a malicious
peer without being detected.  The state of the chaincode enclave itself
remains unchanged after initialization and is stored by the peer.  However,
all correct peers verify that they only interact with chaincode enclaves
registered on the ledger itself.  The ledger enclave may also contain an
encryption key for protecting data on the ledger through the chaincode
library, which handles state encryption and decryption transparently.

\section{Evaluation}\label{sec:eval}
\label{sec:evaluation}

We have built a prototype of the design for secure chaincode executing
using Intel~\ac{SGX} with Hyperledger Fabric.  This section describes the
implementation and reports on the evaluation of the \ac{SGX} prototype for
the blockchain auction application.

\subsection{Implementation}\label{sec:impl}

We implemented the prototype on top of Hyperledger Fabric~1.0.  Each
component of the architecture in Section~\ref{sec:arch} has been integrated
with the Fabric peer code.  We use the Intel~\ac{SGX}~SDK for
Linux~2.1.2~\cite{sgx-sdk-linux} to implement the components residing in an
enclave such as the chaincode library and the chaincode.  These components
are written in C/C++.  The other components, such as the untrusted part of
the chaincode enclave, the enclave registry, and the enclave transaction
validator are written in Go.

The ledger enclave runs as a system chaincode, which allows it to be
integrated in the peer without major changes.  This means that the
chaincode enclave can access the ledger enclave via
\emph{chaincode-to-chaincode (cc2cc)} invocations.
The ledger enclave uses a simple KVS based on
\emph{std::map} to store the integrity metadata.  

The prototype supports a subset of the original fabric chaincode shim and
provides \op{getState}, \op{getRangeQuery}, and \op{putState} calls.  Those
functions are implemented with the help of the untrusted part of the
chaincode enclave and manage the data in the KVS on persistent storage.
When the chaincode accesses the KVS to retrieve a value, the prototype
contacts the trusted ledger enclave via a cc2cc invocation in order to
retrieve the corresponding integrity metadata.  This integrity protection
uses HMAC-SHA256 with a verification key generated during bootstrapping
(Sec.~\ref{sec:approach:bootstrapping}).

The enclave registry interacts with the \ac{IAS} using REST.  The registry
is complemented by a custom VSCC that runs on all peers and that verifies
the attestation report returned by the \ac{IAS}.
The enclave transaction validator is implemented in the form of a custom
VSCC.  It verifies the SGX-specific signatures on the response for the
transactions produced by the chaincode enclave, and obtains the
corresponding public key from the enclave registry.
Signatures use 256-bit ECDSA, as provided by Go v1.10 and the \ac{SGX}~SDK.
The ledger state as well as the transaction proposals and proposal
responses are encrypted by default with 128-bit AES-GCM.  The AES key for
encrypting the proposal is established using a Diffie-Hellman key
derivation scheme available within~SGX.

\subsection{Auction prototype}

In order to demonstrate and evaluate the overhead of the approach, we have
implemented the blockchain auction mentioned earlier.

The auction chaincode runs in an enclave.  A client, the \emph{auctioneer},
creates a new auction by invoking $\op{create(\var{params})}$ at a peer,
and receives in return a unique auction identifier $\var{auction}$, which
is used for any subsequent interaction with this auction.  The invocation
specifies a name for the auction, a description of the asset, and more.
When creating a new auction, the chaincode accesses the KVS (using
\op{getState}) and ensures that no auction with the same name already
exists.  Then it stores \var{params} using \op{putState} and initializes a
placeholder that will store the bids.  After the auction has been created,
it becomes \emph{active} and remains so until the auctioneer invokes
$\op{close(auction)}$.

While the auction is active, clients acting as \emph{bidders} may submit
encrypted bids to the auction by invoking
$\op{bid(\var{val}, \var{auction})}$, where $\var{val}$ denotes the value
the bidder wants to offer for the asset.
Each bid is stored on the blockchain as a tuple of the form $(\var{key},
\var{val}) = (\var{auction}.\var{client}, \var{val})$.

The auctioneer may close the auction at an arbitrary time by invoking
$\op{close(\var{auction})}$.  This transaction acts as the barrier
described in Section~\ref{sec:problem} and writes the updated auction
status to the blockchain using \op{putState}.  Once the auctioneer sees
from its ledger that the auction is closed, it invokes
$\op{evaluate(\var{auction}})$.  When the chaincode enclave receives this
transaction, it determines the bidder with the highest bid and issues the
transfer of the asset in exchange of the value of the bid.

\subsection{Experimental setup}

We deploy a Fabric network with a \emph{solo} ordering service (one trusted
node) using a single channel and three peers.  Each peer and the ordering
service run on a separate Supermicro 5019-MR server with a 3.4GHz four-core
E3-1230 V5 Intel CPU that provides SGX support. All machines are equipped
with 32~GB of memory, 1~Gbps network connection, and a SATA SSD drive; they
run Ubuntu Linux 16.04 LTS Server with the generic 4.13.0-32 Linux kernel.
For reporting transaction throughput we use an increasing number of clients
build with the Fabric Client SDK for
Go~(\url{https://github.com/hyperledger/fabric-sdk-go}) invoking concurrent
transaction over a period of at least 30 seconds and report the average.

As a \emph{baseline} for our experiments we run the same auction chaincode
written in Go in an unprotected environment, executed by an unmodified
Fabric peer.  For comparison this chaincode also uses 128-bit AES-GCM
encryption to seal bids and to encrypt the auction state.  However, since
the peer knows the key, this does not hide the auction data from the peer.

\subsection{Measurements}

\paragraph{TCB Size.}
The trusted computing base (TCB) of our prototype includes the chaincode
enclave and the ledger enclave; all other components of the peer are
considered to be untrusted.  Taken together, the system consists of
approximately 5,000 lines of trusted C/C++ code and 4,000 lines of
untrusted C/C++ and Go code.  The trusted ledger enclave makes up the
majority of the code base, with roughly 3,800 lines, whereas the chaincode
enclave only comprises about 1,200 lines.  The auction chaincode itself is
200 lines of C/C++.  In contrast to a solution where the entire Fabric peer
($\geq 100,000$ lines) or the entire Linux kernel ($\geq 25$M lines) is
executed in the trusted environment, our approach clearly fulfills the goal
of minimizing the TCB.  This facilitates its security analysis through
code reviews or automated verification.

\paragraph{Transaction size.}

In a preliminary experiment we evaluated the transaction sizes for the
auction transactions.  We observed an average transaction size of 3kB for
$\op{bid()}$ and 3.5kB for $\op{evaluate()}$ (with a readset containing 10
bids).  The transactions contain a constant overhead of about 100B for a
256-bit ECDSA signature that is formatted with JSON and produced by the
chaincode enclave as described in Section~\ref{sec:approach}.  The overhead
introduced by our approach in relation to the given transaction size in
Fabric remains relatively small.  As also reported in~\cite{fabric18},
transaction in Fabric are large because they contain PEM-encoded
certificates.

\paragraph{Transaction endorsement.}

Next we study the endorsement throughput and latency of our approach with an
increasing the number of clients.  In this workload we use up to 128 clients,
which concurrently invoke transactions at a single endorsement peer.  Each
client invokes $\op{noop}$ and $\op{submit}$ transactions in a closed loop,
respectively.
The auction chaincode returns immediately for a $\op{noop}$.
The $\op{submit}$ transaction, on the other hand,
receives a bid, encrypts it, and stores it on the blockchain.  We
compare the auction chaincode executed in the SGX enclave with the native,
unprotected execution.  

\begin{figure}[ht!]
    \centering
    \vspace*{-3mm}
    \includegraphics[width=8cm]{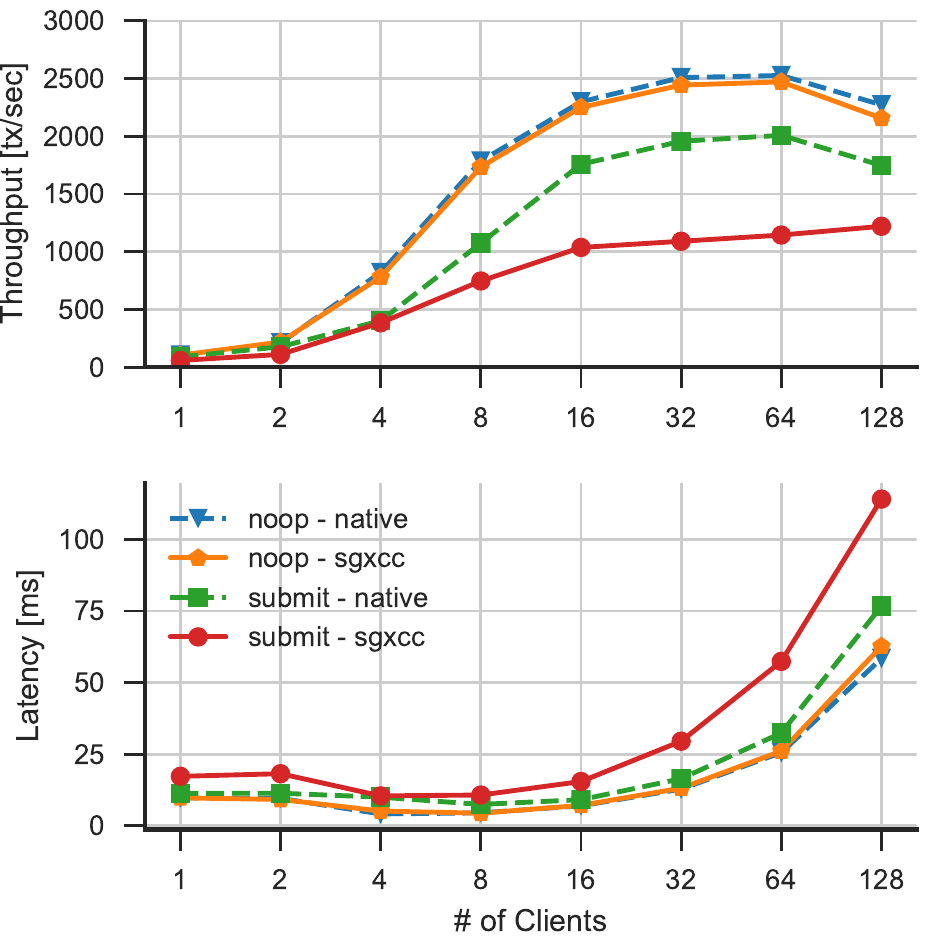}
    \vspace*{-4mm}
    \caption{Endorsement throughput with different numbers of clients.}
    \label{fig:eval:endorsement}
    \vspace*{-3mm}
\end{figure}

As Fig.~\ref{fig:eval:endorsement} shows, the throughput and the latency of
the $\op{noop}$ transaction behave almost identically.  The SGX-based
approach follows the
baseline and scales almost linearly until reaching saturation at 16 clients.
We observe that for the $\op{noop}$ transaction our approach reaches
0.93x--0.99x the throughput of the native execution. On the other hand, when
executing $\op{submit}$ transactions execution in SGX reaches 0.55x--95x the
throughput of the native execution.  In particular, before saturation our
approach shows an increased latency of about 6ms compared to the native
execution.  The reason is that the auction chaincode invokes \op{getState} to
retrieve data from the KVS and additionally fetches the corresponding
integrity metadata from the ledger enclave using cc2cc invocations.
We profiled the response latency and present the breakdown in
Table~\ref{tab:eval:tcb}.  The table shows that retrieving data from the KVS
takes 0.37ms, whereas cc2cc invocation takes 2.59ms to return.  The actual
invocation of the ledger enclave takes less than a millisecond, thus the
majority of the response time is spent for the communication between the two
chaincodes.
In this experiment we see that cc2cc invocations are relatively expensive.
Our choice of implementing the ledger enclave as a chaincode allowed for a
simple integration and coupling with the chaincode enclave, but it comes
with noticeable overhead.  This overhead can be reduced by moving the
ledger enclave into the peer itself, so that it directly provides
Fabric's chaincode API.

\begin{table}
  \small
    \centering

    \begin{tabular}{l l l l l}
        \toprule
        & mean & $\sigma_t$ & $t_\var{min}$ & $t_\var{max}$ \\
        \midrule
		Decrypt tx & 0.20 & 0.04 & 0.17 & 0.38 \\
		\op{getState} & 0.37 & 0.23 & 0.12 & 4.54 \\
		\op{cc2cc} & 2.59 & 1.42 & 1.08 & 11.44 \\
		Ledger enclave  & 0.68 & 0.16 & 0.52 & 1.42 \\
		Decrypt \& verify state & 0.06 & 0.02 & 0.04 & 0.38 \\
		Sign response & 0.226 & 0.045 & 0.179 & 0.402 \\

        \bottomrule
    \end{tabular}

    \vspace*{1mm}
    \caption{Endorsement latency breakdown for \op{submit} transaction with 4 clients showing
the average response time in milliseconds.}
\label{tab:eval:tcb}
    \vspace*{-5mm}
\end{table}

\paragraph{Auction evaluation.}

\begin{figure}[ht!]
    \centering
    \vspace*{-3mm}
    \includegraphics[width=8cm]{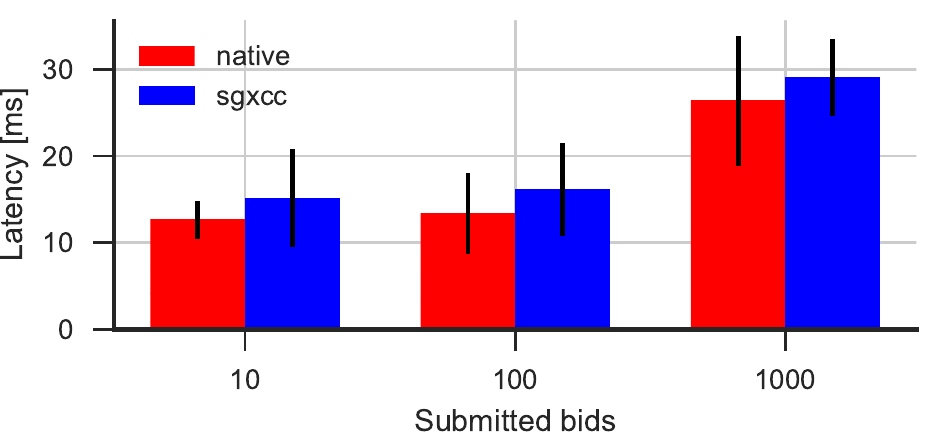}
    \vspace*{-4mm}
    \caption{Auction evaluation latency with different number of submitted bids.}
    \label{fig:eval:range_queries}
    \vspace*{-3mm}
\end{figure}

We also investigate the performance of range queries used by the auction
chaincode to read all submitted bids for determining the winner.  We
measure the response latency for the $\op{evaluate}$ transaction for
different numbers of submitted bids, shown in
Fig.~\ref{fig:eval:range_queries}.  As expected, we observe that latency
increases with larger number of submitted bids.  The relative overhead of
executing in an SGX enclave remains constant for a small number of
submitted bids, at about ~20\%, and for larger numbers it decreases to
10\%.  This experiment shows that using range queries reduces the number of
cc2cc invocations, which improves the performance.

\paragraph{End-to-End performance.}

\begin{figure}[ht!]
    \centering

    \includegraphics[width=8cm]{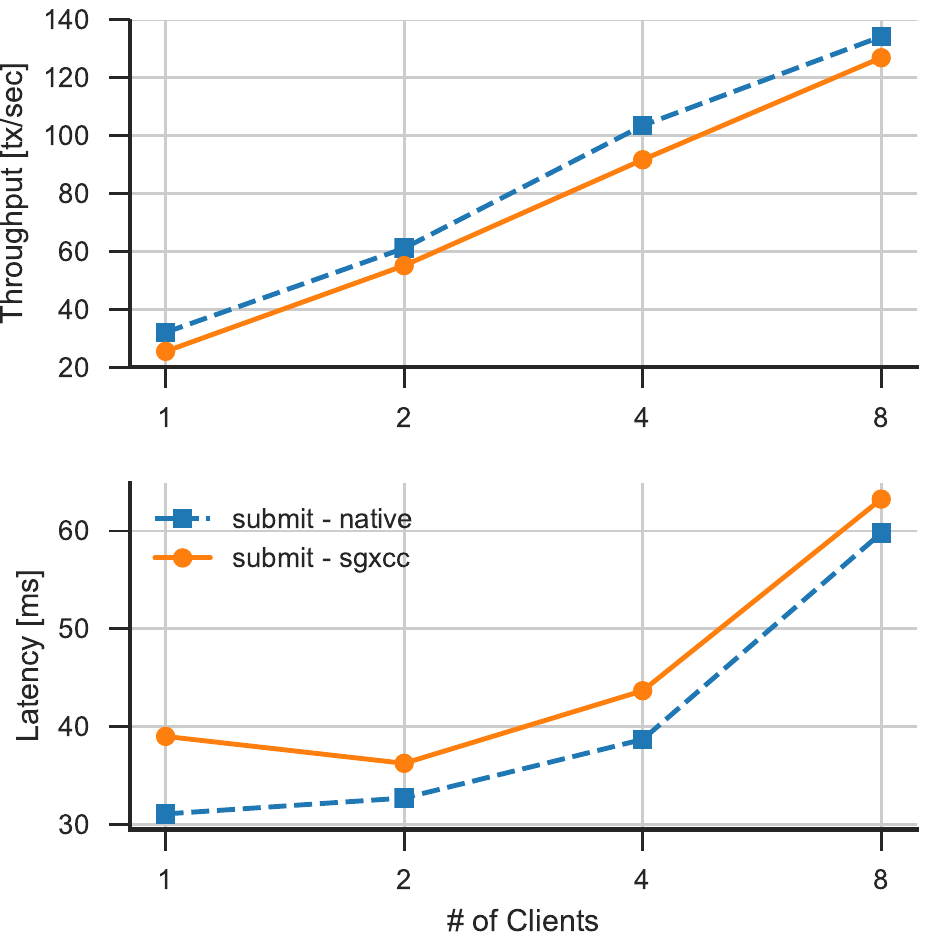}
    \vspace*{-4mm}
    \caption{End-to-end throughput with different numbers of clients.}
    \label{fig:eval:end2end}
    \vspace*{-5mm}
\end{figure}

In the last experiment we study overall end-to-end response latency for the
auction.  The workload is the same as in the transaction endorsement experiment but here
we measure the throughput and latency for the entire transaction flow
including endorsement, ordering, and validation.  
We use the default block size configuration with 10 transactions per block. We observe that execution in SGX reaches 0.80x–0.95x of the throughput achieved by the native execution, as shown in Fig.~\ref{fig:eval:end2end}.

\section{Related Work}
\label{sec:related}

Trusted execution technology is envisaged to play an important role in the
context of blockchains, especially for enterprise applications and in
consortium blockchains.  The two most prominent uses of TEEs are to execute
smart contracts for keeping data private and to represent off-chain data
securely.  We review these two and further related applications of TEEs in
this section.

\paragraph{Smart-contract execution with Intel SGX\@.}
Several approaches have recently suggested to execute blockchain
applications and smart contracts within Intel~SGX\@.

The most prominent among these and most closely related to our work is the
\emph{Coco framework}, announced by Microsoft in a white
paper~\cite{coco-whitepaper}.  It provides a set of building blocks based
on Intel~\ac{SGX} that can be used to secure blockchain systems. Coco
integrates consensus algorithms, distributed ledger state, and a runtime
environment for executing smart-contract transactions in SGX enclaves.  It
appears to be derived from Ethereum and mentions the Ethereum Virtual
Machine (EVM) for its core data structures and protocols. The components of
Coco are described as potentially separate enclaves, but conceptually the
entire blockchain node (corresponding to a peer in Fabric) resides in
\ac{SGX}.  Coco exploits concepts from a proposed related blockchain system
called \emph{VOLT}~\cite{msr-volt}.  Since only a white paper for Coco is
available, however, it is difficult to clearly assess the framework.
It is clear that Coco suffers from a large TCB with its problems as
discussed in Section~\ref{sec:strawman}, including potential malicious
interactions across smart contracts.  Our approach isolates each
application within its own enclave with the minimal amount of code
necessary and follows the philosophy of minimizing the~TCB.

The \emph{R3 Corda} distributed ledger platform has also announced a
privacy feature using~\ac{SGX} in a white paper~\cite{corda-whitepaper}.
In Corda, some aspects of the transaction validation are envisaged to take
place inside an \ac{SGX} enclave, potentially running at untrusted nodes.
By executing the transaction validation in an enclave, the identities
involved in a transaction can be encrypted on the blockchain and are only
revealed inside an enclave during validation.  Corda strives to port a Java
runtime environment (JRE) in order to execute native Corda smart contracts
within an enclave.  Compared to our approach, this introduces some runtime
overhead for the JRE and an increased attack surface with the larger TCB,
but it benefits from portability of the applications written in the same
language.

The \emph{IBM Blockchain Platform}~\cite{ibm-ibp} offers an enterprise
blockchain-as-a-service solution allowing for deployment of a Fabric
network using Secure Service Container~(SSC) technology~\cite{ibm-ssc} on
\emph{IBM Z} systems.  The platform runs the whole peer and its Linux
operating system within a secure container, which is shielded from access
of the host and host administrator, similar to SGX.  This means that
running Fabric within a Secure Service Container requires specialized
mainframe hardware and has a large TCB.  In contrast, our approach for
running Fabric with SGX works with commodity systems and minimizes the TCB
for each smart-contract application.

Several academic papers have also suggested to run smart contracts inside
SGX for confidentiality, e.g., in the ``Ring of Gyges''~\cite{JKS16} or in
Hawk~\cite{KMSWP16}.

\paragraph{Blockchain oracles and off-chain data.}
Other works~\cite{bletchley-whitepaper,ZCCJS16} realize trusted ``oracles''
for blockchain smart contracts using \ac{SGX}.  Oracles are data feeds
external to the blockchain that inform a smart contract about ``facts'' in
the environment.  They extend the scope of inputs to which an application
can respond and serve as trusted sources and triggers for actions on the
blockchain.  Leveraging trusted execution technology enhances the
trustworthiness of an oracle and allows to verify the correctness of the
data source.  This work is orthogonal to our approach, which could also
benefit from oracles that exploit trusted hardware.

Teechain~\cite{teechain2017} is a system to perform off-chain payments on
top of Bitcoin.  It leverages~\ac{SGX} to establish stateful payment
channels among mistrusting parties.  Such off-chain channels resolve
exchanges bilaterally without incurring a blockchain transaction for every
exchange, in the normal case when both parties are honest.  Payment
channels expected to boost the overall throughput of a blockchain-based
payment system.  Through the use of \ac{SGX}, Teechain relaxes the
synchrony assumptions in existing payment channels and gains efficiency.

\paragraph{Consensus protocols.}

Another line of work leverages trusted execution to enhance the resilience
and performance of consensus protocols.  Based on traditional BFT
protocols, systems such as \emph{TrInc}~\cite{levin2009},
\emph{CheapBFT}~\cite{kbcdkm12}, or Hybster~\cite{behl17hybster} have shown how to
enhance state-machine replication with trusted specialized hardware
devices, FPGAs, and SGX enclaves, respectively.

Some blockchain-specific peer-to-peer consensus protocols have been
introduced proposed that scale to a large number of nodes based on trusted
hardware.  \emph{REM}~\cite{ZEEJR17}, for example, introduces
\emph{Proofs-of-Useful-Work} that are run within \ac{SGX} in order to reach
consensus on a public blockchain.
 
Furthermore, the consensus model of the Hyperledger Sawtooth platform
(\url{https://github.com/hyperledger/sawtooth-core}), originally
contributed by Intel, includes the \emph{Proof-of-Elapsed Time (PoET)}
consensus protocol based on \ac{SGX}.  It replaces the proof-of-work
function for leader election in Bitcoin's Nakamoto consensus with a
mandatory, random waiting time imposed by an enclave running in \ac{SGX};
otherwise the protocol is similar to the Nakamoto consensus.

\paragraph{State continuity and TEEs.}

Kaptchuk et al.~\cite{KMG17} address state continuity for memoryless secure
processors that have access to a distributed ledger.  They construct a
generic protocol for detecting rollback attacks, assuming the processor is
always given access to latest ledger state.  This is an interesting
conceptual approach to detect rollbacks, assuming an idealized trusted
ledger that cannot be rolled back.

\section{Conclusion}\label{sec:conclusion}

This work has explored some pitfalls that arise from the combination of
trusted execution with blockchains.  In particular,
smart-contract execution with Intel SGX promises protection for blockchain
applications with strong privacy demands.  However, since enclaves are
susceptible to rollback attacks, a malicious blockchain peer may break
confidentiality by resetting an enclave to manipulate the operation
invocation order.
We have presented a solution that selectively utilizes SGX,
minimizes the TCB, and handles 
rollbacks for the Hyperledger Fabric platform. 
An evaluation has shown that the overhead of our approach is within
10\%–20\% for a sealed-bid auction application.

\bibliographystyle{ACM-Reference-Format}
\bibliography{ref}

\end{document}